\documentclass[a4paper,10pt]{article}
\usepackage{amssymb,amscd,amsmath,amsfonts,graphicx}

\newtheorem{proposition}{Lemma}
\newtheorem{theorem}{Theorem}
\newtheorem{cor}{Corollary}

\def\mc{\mathcal}
 \def\mb{\mathbb}

\def\disp{\displaystyle}
\begin{document}
\centerline{\large{Reflected backward stochastic differential equations}}
 \centerline{\large{ and a class of non linear dynamic pricing rule}}

\vspace{0.5cm}
\centerline{Marie-Amelie Morlais, ETH Zurich, Switzerland.}
\centerline{Ramistrasse 101, 8006 Zurich,} 
\centerline{Tel: +41 44 632 5859, e-mail: marieamelie.morlais@free.fr
}
\vspace{0.5cm}

\begin{abstract}
In that paper, we provide a new characterization of the solutions of specific reflected backward stochastic differential equations (or RBSDEs) whose driver $g$ is convex and has quadratic growth in its second variable: this is done by introducing the extended notion of $g$-Snell enveloppe. Then, in a second step, we relate this representation to a specific class of dynamic monetary concave functionals already introduced in a discrete time setting. This connection implies that the solution, characterized by means of non linear expectations, has again the time consistency property. 
\end{abstract}
\newpage
\section{Motivation}
In that paper, we consider a specific form of reflected backward stochastic differential equations (RBSDEs in short)
which are defined on a finite time horizon $T$. Some particular RBSDEs are studied, for instance, in \cite{ElkarouiKapoudian} in connection with PDE obstacle problems or also in \cite{ElkarQuenez}, in connection with the problem of pricing an American contingent claim. We consider here, in a brownian setting, a class of reflected BSDEs (those can be viewed as a kind of generalized BSDEs). In this brownian setting and denoting by $\mc{F}$ the brownian filtration, a solution of the RBSDE with parameters ($g$, $B$, $U$) is a triple of $\mc{F}$-adapted processes ($Y, Z, K$). 
In all that paper, the notation $g$ refers to the driver, $B$ refers to the terminal condition (this is a $\mc{F}_{T}$-measurable random variable) and the process $U$ refers to the upper constraint. Compared with usual BSDEs, the difference is the presence of an additional constraint on the solution: this implies the presence of the increasing process $K$, whose aim is to force the solution to satisfy this constraint.\\

Our objective is twofolds: we first characterize, by means of non linear expectations, the unique solution of a class of RBSDEs having a driver with quadratic growth.
 Under this last condition on the driver, existence for minimal and maximal solutions
 of such RBSDEs has been established in \cite{Kobyletal}: in the aforementionned paper, the authors refer to the results and methods employed in \cite{mkobylanski}: the originality of the present study is not the result in itself but consists rather in proving the existence of a non linear Doob-Meyer's decomposition. As in \cite{Kobyletal}, we rely on already established existence, uniqueness and comparison results for solution of quadratic BSDEs and also on fine properties of the driver $g$. In a second step and using the characterization obtained, we provide a connection with one specific dynamic concave utility functional (also denoted by DMCUF in the sequel) or equivalently, up to a minus sign, with one dynamic convex risk functional: this specific DMCUF is discussed in a discrete time setting in \cite{cheriditodelbaen} and in a continuous setting in the more recent study \cite{bionnadal2007}: in the first aforementionned reference, it is constructed from a given time consistent DMCUF and it is proved, in particular, that the extension is again time consistent. Besides, this construction extends to the non subadditive case the pricing rule introduced in \cite{ElkarQuenez}. Some other major references are \cite{KloppelSchweizer}, \cite{Barrieuelkar}, \cite{RosazzaG} and \cite{Peng2003}. In \cite{KloppelSchweizer}, the authors study some properties of these DMCUF, especially the inf-convolution procedure of these functionals and they provide links with utility indifference valuation. The se-cond reference \cite{Barrieuelkar} deals
with a general review of the links between dynamic risk measures and BSDEs and they look at their respective
 properties and representations: this is done by studying the connection with hedging problems of interest in finance. The two last papers give further analysis of both conditional dynamic
risk measures and solutions of some particular BSDEs, the so-called conditional
 $g$-expectations.\\

The present paper is structured as follows: in a first section, we give preliminary notations and results about quadratic BSDEs and we introduce the specific class of RBSDEs we are interested in. Then, to characterize the solution of these RBSDEs, we prove the existence of an extended decomposition of Doob-Meyer's type for non linear expectations, which is the main ingredient to achieve the representation of the solution. The last section provides both the connection with one specific DMCUF and the link with the forward
price for the American claim obtained via utility maximization (analogously to \cite{ElKaretRouge}).

\section{Theoretical study of the quadratic RBSDE}
\subsection{Notations and preliminaries}
We consider a probability space ($\Omega, \mb{F}, \mb{P}$), on which is defined a $d$-dimensional brownian motion $W$ and we denote by $\mc{F}$ the natural filtration generated by $W$ and completed by $\mc{N}$ consisting in all the $\mb{P}$-null sets. 
The form of the quadratic RBSDE we are interested in is given as follows
\[(\textrm{Eq2.1}) \quad\left\{ \quad  \begin{array}{l}
 (i) \; \forall \; t, \quad \disp{Y_{t} = B + \int_{t}^{T}f_{0}(s, Z_{s})ds -(K_{T} - K_{t}) -\int_{t}^{T}Z_{s}dW_{s},\quad  }\\
 K \; \textrm{is increasing and s.t.} \;\\
(ii)\; \disp{\int_{0}^{T} \big(Y_{s} - U_{s} \big) dK_{s} = 0}.  \\ 
   
(iii) \; \forall \; t, \quad Y_{t} \le U_{t}.\quad  \;\\
\end{array} \right. 
\]
\indent A solution of the RBSDE is a triple ($Y, \;Z, \; K$) satisfying ($\textrm{Eq2.1}$) such that $(Y, Z)$ is a pair of adapted processes in $S^{\infty} \times \mc{H}^{2}$ and $K$ is an increasing adapted process.
$S^{\infty}$ denotes the set of all the continuous processes $Y$ such that $\textrm{ess} \disp{\sup_{\omega, s}|Y_{s}| }  < \infty$ and $\mc{H}^{2}$ denotes the set of all the progressively measurable processes $Z$ such that $\disp{\mb{E}(\int_{0}^{T}|Z_{s}|^{2}ds) < \infty}$. In all that paper, $Z \cdot W$ will denote the stochastic integral of $Z$ with respect to $W$.
The process, denoted by $U$ in ($\textrm{Eq2.1}$), which stands for the upper barrier of any solution of the RBSDE, is assumed to be in $S^{\infty}$. To ensure the well-posedness of the problem, we also need to have: $B \le U_{T}, \; \mb{P}\textrm{-a.s.} $\\
In the sequel, $B$ is a bounded $\mc{F}_{T}$-measurable random variable and the driver $f_{0}$ satisfies the standing assumptions ($H_{0}$) and $(H_{1})$
\[(H_{0}) \;\left\{ \; \begin{array}{l}
0 \le  f_{0}(s, z) \le C\big(1 + |z|^{2}\big),\\
\forall \; t, \; \;\left(\disp{\int_{0}^{t}f_{0}(s, 0)ds}\right) \in L^{\infty}(\mc{F}_ {t})\\ 
f_{0} \; \textrm{is convex w.r.t.} \; z,\\
f_{0} \; \textrm{is independent of}\; y.\\
                                       \end{array} \right.
\]

$$(H_{1}) \quad \exists \; \kappa \in BMO(W) \; \forall \;z, \; z^{'} \; \; \frac{|f_{0}(s,z) - f_{0}(s, z^{'})|}{|z - z^{'}|} \le C(\kappa + \;|z| + \;|z^{'}| ). $$
 This last BMO property \footnote{This condition on the increments of the driver w.r.t $z$ is analogous to the one given in \cite{Imkellerethu}.} stated in ($H_{1}$) is crucial in the proof of the uniqueness result we provide in Section 2.3.
We now introduce the normalized driver $g$ 
\begin{equation}\label{eq: expressiong}
g(s, z):= f_{0}(s, z) - f_{0}(s, 0),
\end{equation}
 which is such that: $g(s, 0) \equiv 0 $, and for later use, we introduce the notation $\mc{E}_{g}(B|\mc{F}_{t})$ for the unique process $Y$ satisfying 
\begin{equation}\label{eq: BSDEdriverg} Y_{t} - B := \disp{\int_{t}^{T}g(s, Z_{s})ds - \int_{t}^{T}Z_{s}dW_{s}},\end{equation}
which is a BSDE with driver $g$ and terminal condition $B$.
This process corresponds to the conditional non linear expectation (defined in \cite{Peng2003}) which has been introduced for a driver $g$ such that $g =g(t, z)$ and $g$ is lipschitz w.r.t the variable $z$.\\
\indent 
Here, using both assumption ($H_{0}$) on $f_{0}$ and the results on quadratic BSDEs obtained in \cite{mkobylanski}, we can extend this notion of non linear expectation to the case of a quadratic driver $g$ defined such as in (\ref{eq: expressiong}). Furthermore, we check that it satisfies the same properties as the (conditional) $g$-expectation introduced in \cite{Peng2003}\\
$\bullet \;$ it is translation invariant  
$$\forall\; \xi\in L^{\infty}(\mc{F}_{T}), \; \eta \in L^{\infty}(\mc{F}_{t}), \quad \mc{E}_{g}\big(\xi + \eta |\mc{F}_{t}\big) =\mc{E}_{g}\big(\xi |\mc{F}_{t}\big)  + \eta. $$
$\bullet \;$ it is monotone
$$ \forall \; \xi\in L^{\infty}(\mc{F}_{T}), \; \eta \in L^{\infty}(\mc{F}_{T}), \quad \xi \le \eta \; \; \Rightarrow (\mc{E}_{g}\big(\xi |\mc{F}_{t}\big) \le \mc{E}_{g}\big(\eta |\mc{F}_{t}\big) )$$
$\bullet \;$ it is constant preserving
$$ \forall \eta \in L^{\infty}(\mc{F}_{t}), \quad \mc{E}_{g}\big(\eta |\mc{F}_{t}\big) = \eta.$$
$\bullet \;$ it has the strong time consistence property
 $$ \forall \; \xi \in L^{\infty}(\mc{F}_{T}),\forall \; t \le s, \quad \mc{E}_{g}(B | \mc{F}_{t}) =\mc{E}_{g}(\mc{E}_{g}(B | \mc{F}_{s}) | \mc{F}_{t}). $$
The invariance by translation property results from the $y$-independence of $f_{0}$, the monotonicity comes from the comparison result for quadratic BSDEs and the constant preserving property results from the fact that $g(s, 0) \equiv 0$. The last property is a standard one, which is satisfied by any solution of the BSDE given by (\ref{eq: BSDEdriverg}).\\
\subsubsection*{Comments}
$\bullet $ Some connections between properties of the driver and those of the related conditional $g$-expectation have been established in \cite{Briandhucoquet} in the case of particular $g$ expectations called dominated $g$ expectations. In particular, the convexity property of the driver entails that the $g$-expectation is itself convex. This last property is meaningful considering the connection with finance: indeed, a proper conditional $g$-expectation (i.e. satisfying the four aforementionned properties) is related to a conditional risk measure via: $\rho_{t}^{g}(\xi) :=\mc{E}_{g}(-\xi |\mc{F}_{t})$. The financial interpretation of the convexity property is that diversification in portfolio choice reduces the risk assessed through the risk measure. \\
$ \bullet $ A largely used example of non linear expectation is provided by the choice of the quadratic function $g_{\alpha}(s, z):= \frac{\alpha}{2}|z|^{2}$. It is well known that the unique solution of the BSDE($g_{\alpha},B$) is
$$ \mc{E}_{g_{\alpha}}\big(B |\mc{F}_{t} \big) = \frac{1}{\alpha}\ln \big(\mb{E}(e^{\alpha B} |\mc{F}_{t}) \big),$$ 
 and this is linked to the conditional entropic risk measure via the formula
$$\forall \; t \in [0, T], \quad \rho_{t}^{\alpha}(B) := \mc{E}_{g_{\alpha}}\big(-B |\mc{F}_{t} \big). $$ 
\subsection{The main result}

\begin{theorem}\label{SolutionRBSDE}
 Let ($Y,\; Z,\; K $) be a solution of the RBSDE then it satisfies
$$ Y_{t} := \textrm{ess} \disp{\inf_{\tau \in \mc{S}_{t, T}}\mc{E}_{g}\left(B\mathbf{1}_{\tau = T} +U_{\tau}\mathbf{1}_{\tau < T} +\int_{t}^{\tau}f_{0}(s, 0)ds |\mc{F}_{t} \right) },$$
where $g(s, z) := f_{0}(s, z) - f_{0}(s, 0)$ and $\mc{S}_{t}$ stands for the set of all the stopping times taking their values in $[t, T]$.\\
Besides, the process $\tilde{Y} :=\disp{ Y + \int_{0}^{\cdot}f_{0}(s, 0)ds}$ is the greatest $g$-submartingale for which $Y$ solves the RBSDE: i.e., if there exists
another $g$-submartingale $\tilde{Y}^{'} := Y^{'} + \disp{\int_{0}^{\cdot}f_{0}(s, 0)ds}$ with $Y^{'}$ satisfying equation (i) in ($\textrm{Eq2.1}$) and such that: $Y^{'} \le U $, then $\tilde{Y}^{'}$ satisfies: $\tilde{Y}^{'} \le \tilde{Y}$.
\end{theorem}
Before justifying Theorem \ref{SolutionRBSDE}, which characterizes the unique solution of ($\textrm{Eq2.1}$), we provide some preliminary results: a major part of the proofs is standard but, to make the presentation of this paper self contained, we give in next subsection the outline of the proofs adapted to our setting.\\


\subsection{Auxiliary results on quadratic BSDEs}
In this part and for later use, we provide some major existence, uniqueness and comparison results for BSDEs with a driver satisfying $(H_{0})$ and $(H_{1})$.
We denote here by BSDE($f_{0}, \; B $) the equation given by
$$ \quad Y_{t} = B +\disp{ \int_{t}^{T}f_{0}(s, Z_{s})ds - \int_{t}^{T} Z_{s}dW_{s}.}$$
\subsubsection*{Statement of the main results}

\begin{theorem}\label{existetuniq}
 Under assumptions ($H_{0}$) and ($H_{1}$) on $f_{0}$ and as soon as $B$ is bounded, the BSDE($f_{0}, \; B $) has a unique solution in $\mc{S}^{\infty} \times \mc{H}^{2}$.
\end{theorem}

\begin{cor}\label{comparison}
 Let $f$ and $f^{'}$ be two generators satisfying both ($H_{0}$) and ($H_{1}$) and let $B$, $B^{'}$ be two bounded $\mc{F}_{T}$-measurable random variables. If besides, we have
$$ f \le f^{'} \quad \textrm{and} \quad B \le B^{'},$$
then, the respective solutions ($Y, Z$) and ($Y^{'}, Z^{'}$) of the BSDEs given by ($f, \; B $) and ($f^{'}, \; B^{'}$) satisfy
$$ \mb{P}\textrm{-a.s. and for all}\; s, \quad Y_{s} \leq Y_{s}^{'}.  $$
\end{cor}

For later use, we reestablish standard a priori estimates for any solution of a BSDE with driver satisfying ($H_{0}$). Some of the arguments and methods have already been used in \cite{Briandethu}.
\begin{proposition}\label{aprioriestimates}
Considering a BSDE with parameters ($f,\; B$) with its driver $f$ satisfying $(H_{0})$ and its terminal condition $B$ bounded, there exists estimates depending only on $C$, $T$ ($C$ is given in ($H_{0}$) and on the terminal condition $B$ such that, for any solution ($Y, Z$) in $S^{\infty} \times \mc{H}^{2}$ of the BSDE($f,\; B$) and for any stopping time $\tau$, we have   \\
(i)
If the process $\tilde{Y}$ is defined by
\begin{equation}\label{eq: correspondence}
 \tilde{Y}_{t}:= Y_{t} + \disp{\int_{0}^{t}f_{0}(s, 0)ds }  \; \;\left(\textrm{and:} \; \tilde{B} := B + \disp{\int_{0}^{T}f_{0}(0, s)ds}\right),
\end{equation} then, 
$$\mb{P}\textrm{-a.s. and for all} \; t, \; \;  \mb{E}\big(\tilde{B} |\mc{F}_{t}\big)  \le \tilde{Y}_{t} \le \frac{1}{C}\ln\left(\mb{E}(\exp(C\tilde{B})|\mc{F}_{t})\right) . $$
(ii) $Z$ satisfies
\\$$\; \quad |Z|_{\textrm{BMO}(W)}:= \disp{\sup_{\tau}\mb{E}\left(\int_{\tau}^{T}|Z_{s}|^{2}ds|\mc{F}_{\tau}\right) \le C^{'}},$$
 with $C^{'}$ depending only on $T$, $C$ and $|Y|_{S^{\infty}}$.
\end{proposition}

\subsubsection*{A one-to-one correspondence result}
In this step, we prove
the one-to-one correspondence between the solution of the BSDE($f_{0}, \; B$) and the solution of the BSDE($g, \tilde{B}$),
 with driver $g$ given by (\ref{eq: expressiong})
and terminal condition $ \tilde{B} = B + \int_{0}^{T}f_{0}(s, 0)ds$. 
%
If we define $Y$, for all $t$ by
\begin{equation}\label{eq: relationeq} Y_{t} := \mc{E}_{g}\left(B + \int_{t}^{T}f_{0}(s, 0)| \mc{F}_{t}\right),
 \end{equation}
this process solves the BSDE($f_{0}, B$). Besides and as soon as $\disp{\int_{0}^{T}f_{0}(s, 0)ds}$ is in $ S^{\infty}$, $Y$ is itself in $S^{\infty}$. 
Considering now the process $\tilde{Y}$ defined by (\ref{eq: correspondence}) in terms of $Y$ (in Lemma \ref{aprioriestimates}),
both the equality (\ref{eq: relationeq}) and the definition of the $g$-expectation imply that $\tilde{Y}$ solves the BSDE($g, \tilde{B}$) with $\tilde{B}:= B +\disp{\int_{0}^{T}f_{0}(s, 0)ds }$, or equivalently: $\tilde{Y}_{t} = \mc{E}_{g}(\tilde{B} | \mc{F}_{t}).$ Conversely, if $\tilde{Y}$ solves the BSDE($g, \tilde{B}$), then, thanks to the tranlation by invariance property of the conditional $g$-expectation, it follows 
\[ \begin{array}{ll}
 \tilde{Y}_{t} &= \mc{E}_{g}\left( B + \disp{\int_{0}^{t}f_{0}(s, 0)ds + \int_{t}^{T}f_{0}(s, 0)ds} \big| \mc{F}_{t}\right)\\
  \\
  & = \disp{\int_{0}^{t}f_{0}(s, 0)ds} +\mc{E}_{g}\left( B + \disp{ \int_{t}^{T}f_{0}(s, 0)ds} \big| \mc{F}_{t}\right),\\
 \end{array} \]
and hence, equality (\ref{eq: correspondence}) provides the desired one-to-one correspondence result.\\
\indent

\subsubsection*{Outline of the proofs}
\textbf{Proof of Lemma \ref{aprioriestimates}}
We assume here the existence of a solution ($Y, \;Z$) in $S^{\infty} \times \mc{H}^{2}$ of the BSDE with parameters ($f, \; B$).
Relying on the one-to-one correspondence result (\ref{eq: correspondence}), we get: $\tilde{Y}_{t} : =\mc{E}_{g}(\tilde{B} |\mc{F}_{t})$. Hence, assertion (i) is checked as soon as  
\begin{equation}\label{eq: estimgesperance} \mb{E}\big(\tilde{B} |\mc{F}_{t}\big)\le \mc{E}_{g}(\tilde{B} |\mc{F}_{t}) \le \frac{1}{C}\ln\mb{E}\left(\exp(C\tilde{B}) |\mc{F}_{t} \right).
\end{equation}
Furthermore using that, for all $t$: $ \disp{\int_{0}^{t}f_{0}(s, 0)ds} \in L^{\infty}(\mc{F}_{t})$, and relying on the growth condition in $(H_{0})$, we have
$$ \disp{\left|\int_{t}^{T}f_{0}(s, 0)ds\right| \le C(T -t)}.$$
To justify the first inequality in (\ref{eq: estimgesperance}), we use both the positiveness of $g$ and the comparison theorem provided by Corollary \ref{comparison} to claim that $\tilde{Y} :=\big(\mc{E}_{g}(\tilde{B} |\mc{F}_{t})\big)_{t}$ is greater than the solution of the BSDE with parameters ($0, \;\tilde{B}$), which implies: $\tilde{Y}_{t} \ge \mb{E}\big(\tilde{B} |\mc{F}_{t}\big), \; \mb{P}$-a.s. and for all $s$.\\
The other inequality in (\ref{eq: estimgesperance}) results from the application of It\^o's formula to $e^{C \tilde{Y}} $: this yields the following submartingale property
$$ e^{C \tilde{Y}_{t}} \le \mb{E}\big(e^{C \tilde{B}} |\mc{F}_{t} \big),$$
and hence, (\ref{eq: estimgesperance}) follows.\\
\indent 
To prove the estimate (ii), we proceed analogously as in Corollary 4 in \cite{Briandethu}. Since ($Y, \; Z$) is in $S^{\infty} \times \mc{H}^{2}$, we just apply It\^o's formula to $u(Y)$ by setting: $u(y) =\frac{e^{2Cy} -2Cy -1 }{(2C)^{2}}$, between an arbitrary stopping time $\tau$ and $T$. 
\begin{tabbing}
 $u(Y_{ \tau }) $ \= $ = u(B)+ \disp{\int_{\tau}^{T } u^{'}(Y_{s}) f(s, Z_{s})ds }$\\
\\
\> $ \quad  \disp{- \; \int_{\tau}^{T}u^{'}(Y_{s})Z_{s}dW_{s}}
- \frac{1}{2}\disp{\int_{ \tau}^{T} \big(u^{''}(Y_{s})\big)|Z_{s}|^{2}ds }. $\\
\end{tabbing}
We then take the conditional expectation w.r.t. $\mc{F}_{\tau}$:  since $\disp{\int_{0}^{\cdot}u^{'}(Y_{s})Z_{s}dW_{s}} $ is a true martingale, its expectation is equal to zero. Hence, relying on the relations: $u^{''} - 2C u^{'} \equiv 1$ and: $f(s,z) \le C(1 + |z|)^{2}$  ($f$ satisfying $(H_{0})$), it implies
$$u(Y_{\tau})  \le \mb{E}\left(u(B) + \disp{\int_{ \tau}^{T}C u^{'}(Y_{s})ds}-\frac{1}{2}\disp{\int_{ \tau}^{T} |Z_{s}|^{2}ds \big| \mc{F}_{\tau}}\right), $$
or also: $\mb{E}\left(\frac{1}{2}\disp{\int_{ \tau}^{T} |Z_{s}|^{2}ds \big| \mc{F}_{\tau}}\right) \le \mb{E}\left(\disp{u(|B|)+ \int_{\tau}^{T}C u^{'}(|Y_{s}|)ds \big| \mc{F}_{\tau}}\right)$, from which the result follows.\\
\begin{flushright}
 $\square$
\end{flushright}
\textbf{Remark} $\quad$ Without additional difficulty, we can extend these a priori estimates by adding a linear term in $z$ in the expression of the driver: we assume here that the new normalized driver is: $\tilde{g}(s,z) := g(s,z) + \beta z$, and that: $\disp{\int_{0}^{\cdot}\beta_{s} dW_{s}}$ is a BMO martingale (this notion of BMO martingale can be found in \cite{Kazamaki}) and we then introduce an equivalent measure $\mb{P}^{\beta}$ by setting: $\frac{d\mb{P}^{\beta}}{d\mb{P}} := \mc{E}\big(\beta \cdot W \big)$, where $\mc{E}\big(\beta \cdot W \big) $ stands for the stochastic exponential of $\beta \cdot W$. Under these conditions, the Girsanov's tranform $W^{\beta} :=  W - \disp{\int_{0}^{\cdot}\beta_{s}ds}$ is again a brownian motion under $\mb{P}^{\beta}$. Hence, any solution $(\tilde{Y}, \; \tilde{Z})$ of the BSDE with driver $\tilde{g}$ and terminal condition $B$ satisfies
$$ \tilde{Y}_{t} := B + \disp{\int_{t}^{T}g(s, \tilde{Z}_{s})ds -\int_{t}^{T}\tilde{Z}_{s}dW_{s}^{\beta}},$$
which is a new BSDE with parameters ($g, B$) under $\mb{P}^{\beta}$.
Replacing the standard expectation by $\mb{E}^{\mb{P}^{\beta}}$, $\tilde{Y}$ satisfies the same kind of estimates as in assertion (i) of Lemma \ref{aprioriestimates}. Using now the equivalence between $\mb{P}^{\beta}$ and $\mb{P}$, this process is bounded $\mb{P}^{\beta}$ and $\mb{P}$-a.s. Furthermore and thanks to theorem 3.6 in \cite{Kazamaki}, we have that $ \disp{\int_{t}^{\cdot}\tilde{Z}_{s}dW_{s}^{\beta}}$ is in BMO($\mb{P}^{\beta}$), as soon as $\disp{\int_{t}^{\cdot}\tilde{Z}_{s}dW_{s}} $ is in BMO($\mb{P}$).\\

\subsubsection*{Proof of theorem \ref{existetuniq}}
Referring to \cite{mkobylanski}, the existence result in Theorem \ref{existetuniq} for solutions of the BSDE($f, \;B$) follows from the growth assumption in ($H_{0}$). The uniqueness result relies mainly on assumption ($H_{1}$)
and on a standard linearization procedure (as in the case when the generator is lipschitz w.r.t. $z$). To proceed in the quadratic case and as in \cite{Imkellerethu} and assuming that ($Y^{1}, \; Z^{1}$) and ($Y^{2}, \; Z^{2}$) are two solutions of the BSDE($f, \;B$), 
we apply It\^o's formula to $Y^{1, 2} := Y^{1} - Y^{2}$ between $t$ and $\tau \wedge T$ with an arbitrary stopping time $\tau$ (similarly, $Z^{1, 2}$ stands for $Z^{1} - Z^{2}$)
\begin{tabbing}
$Y_{t}^{1, 2} $ \= $ := Y_{\tau \wedge T}^{1, 2} +\disp{\int_{t}^{\tau \wedge T}\left(f_{0}(s, Z_{s}^{1}) - f_{0}(s, Z_{s}^{2}) \right)ds - \int_{t}^{\tau \wedge T}Z_{s}^{1, 2}dW_{s}.} $\\
\end{tabbing}
We then introduce $\lambda$ such that: $\lambda = ( \lambda_{s}(Z_{s}^{1}, Z_{s}^{2})) $ by setting
\[\left\{  \begin{array}{ll}
     \lambda_{s}(Z_{s}^{1}, Z_{s}^{2}) \;& =  \left(\frac{f_{0}(s,Z_{s}^{1}) - f_{0}(s,Z_{s}^{2})}{Z^{1}_{s} - Z_{s}^{2}}\right),  \; \textrm{if}\; \hat{Z} \neq 0,\\
     \\
    \lambda_{s}(Z_{s}^{1}, Z_{s}^{2}) &  = \; 0, \quad \textrm{otherwise}.\\
    \end{array} \right.
  \]  
Thanks to ($H_{1}$),
$$ \big|f_{0}(s, Z_{s}^{1}) - f_{0}(s, Z_{s}^{2})\big| = |\lambda_{s}(Z_{s}^{1}, Z_{s}^{2})| |Z_{s}^{1, 2}| \le C\big(\kappa + |Z_{s}^{1}| + \;|Z_{s}^{ 2}|\big)|Z_{s}^{1, 2}|.$$
 Referring to Kazamaki's criterion (\cite{Kazamaki}), the stochastic exponential of the continuous BMO martingale $\kappa \cdot W$ is a martingale. Besides, the a priori estimates of Lemma \ref{aprioriestimates} entails the BMO property of both $Z^{1} \cdot W$ and $Z^{2} \cdot W$. Hence, setting: $\frac{d\mb{P}^{\lambda}}{d\mb{P}}:= \mc{E}(\lambda \cdot W)$, we define an equivalent measure denoted by $\mb{P}^{\lambda}$. Using Girsanov's theorem,
 $Y^{1, 2}$ is a local submartingale under $\mb{P}^{\lambda}$: hence, there exists an increasing sequence ($\tau^{m}$) of $\mc{F}$ stopping times converging to $T$, taking their values in $[t,\; T]$ and such that ($Y_{\cdot \wedge \tau^{m}}^{1, 2}$) is a submartingale. This means
$$ Y_{t}^{1, 2} \le \mb{E}^{\mb{P}^{\lambda}}\big(Y_{\tau^{m} \wedge T}^{1, 2} | \mc{F}_{t}\big).$$
Thanks to the boundedness of the sequence $(Y_{\tau^{m} \wedge T}^{1, 2}$), the dominated convergence theorem entails that $Y^{1, 2}$ is a submartingale with terminal value equal to zero. 
Reverting the roles of $Y^{1} $ and $Y^{2} $, we get: $Y^{1, 2} \equiv 0$, which ends the proof.\\
The proof of Corollary \ref{comparison} relies on the same computations, if we apply It\^o's formula to $Y - Y^{'}$.

\subsection{Decomposition of Doob-Meyer's type}
In this paragraph, we establish the existence of a decomposition of Doob-Meyer's type for any $g$-submartingale (or supermartingale) with a  generator $g$ satisfying both ($H_{0}$) and ($H_{1}$) and such that $g$ is normalized, i.e.: $g(s, 0) \equiv 0$.
A process $Y$ is called $g$-submartingale (resp. $g$-supermartingale) if it satisfies
$$ \forall \; \; s \le t, \quad  \mc{E}_{g}(Y_{t} | \mc{F}_{s}) \ge Y_{s} \; \; (\textrm{resp.} \; \mc{E}_{g}(Y_{t} | \mc{F}_{s}) \le Y_{s} ).$$ 
 This result is an extension of the decomposition obtained in theorem 4.3 in \cite{Peng2003} in the case of a dominated $g$-expectation (in the paper \cite{Peng2003}, the notion of domination corresponds to the case of a driver having at most linear growth in $z$). 
In the sequel, $Y$ stands for a given $g$-submartingale with terminal value $Y_{T} = B$. 
The aim of this section is to construct an increasing process $A$ such that $Y - A $ is a $g$-martingale. To this end, we first introduce the sequence of penalized BSDEs with parameters ($g^{n}, \; B $), with $g^{n}$ such that
\begin{equation}\label{eq: generateurapproche}
 g^{n}(s, y, z) := g(s, z) - n\big(y - Y_{s} \big) .\end{equation}
Hence, we have
  $$ |g^{n}(s, y, z)| \le C|z|^{2} + n\big(|y| + |Y|_{S^{\infty}}\big), 
$$
 i.e. $g^{n}$ has linear growth w.r.t. $y$ (it is even $n$-Lipschitz w.r.t $y$) and quadratic growth w.r.t. $z$.
Existence and uniqueness results for such kind of BSDEs are given in \cite{Lepeletsanm}. 
We denote by ($y^{n}, z^{n}$) the unique solution of BSDE($g^{n}, \; B$) which satisfies
$$\disp{ y_{t}^{n}:= B + \int_{t}^{T}\big(g(s, z_{s}^{n}) - n(y_{s}^{n} - Y_{s}) \big)ds - \int z_{s}^{n}dW_{s}.}$$
Besides, it is also proved in \cite{Lepeletsanm} that, for all $n$, $(y^{n}, z^{n})$ is in $ S^{\infty} \times \mc{H}^{2}$.\\
The proof of the existence of the decomposition is divided in three mains steps: those steps consist in following the same scheme than in the proof of Theorem 4.3 in \cite{Peng2003} or also in \cite{Briandhucoquet} (for dominated $g$ expectations). Many computations are standard and, for sake of completeness, we provide the outline of the proofs.\\

\subsubsection*{Step 1: properties of the penalized sequence}
\begin{proposition}
 $Y$ being a $g$-submartingale, the sequence of ($y^{n}, z^{n}$) of solutions of the BSDEs($g^{n}, \; B$) with $g^{n}$ given by (\ref{eq: generateurapproche}) satisfies 
$$\mb{P}\textrm{-a.s. and for all} \; n, \quad  y^{n} \ge y^{n + 1} \ge Y.$$
\end{proposition}

To justify that: $ y^{n }\ge Y$, for all $n$, we also refer to the proof given in Lemma 4.11 in \cite{Peng2003}, which holds for dominated $g$-expectations \footnote{the explanation for this notion, defined in \cite{Peng2003}, is provided at the beginning of Section 2.4 (top of this page)}: the key idea of this proof consists in using both the $g$-submartingale property of $Y$ and the construction of ($y^{n}$)
to show that for any positive $\delta$ and for each $n$, $\{y^{n} \le Y -\delta  \}$ is a $\mb{P}$-null set. Then, the monotonicity property of ($y^{n}$) results from the comparison theorem applied here for the BSDEs given by parameters ($g^{n}, \;Y_{T}$) with quadratic drivers $g^{n}:= g^{n}(s, y, z)$ (for these kind of drivers having linear growth w.r.t. $y$, existence results are provided in \cite{Lepeletsanm}).\\
\begin{flushright}
 $\square$
\end{flushright}

\subsubsection*{Step 2: boundedness of processes}
\indent 
For more convenience, we first introduce the increasing process $A^{n}$ by setting: \\$A_{\cdot}^{n} := \disp{n \int_{0}^{\cdot}(y_{s}^{n} - Y_{s})ds}$. In the sequel, a stochastic integral $Z \cdot W$ is in $\mc{H}^{p}$, if: $\mb{E}\big(\int_{0}^{T}|Z_{s}|^{2}ds \big)^{\frac{p}{2}} <\infty.$ 
 Our aim is to prove the boundedness of ($A_{T}^{n}$) and ($z^{n}$) respectively in $L^{p}(\mc{F}_{T})$ and in $\mc{H}^{p}$ for any $p$, $p> 1$. \\
 Due to the quadratic growth w.r.t. $z$ of the driver $g^{n}$, the arguments of this step differ from \cite{Peng2003}. We rely here on the estimates provided by lemma \ref{aprioriestimates}
 on the sequences ($y^{n}$) and ($z^{n}$) and we follow the same scheme as the one given in \cite{Humapengyao}.
To obtain boundedness of ($A_{T}^{n} $) in $L^{p}(\mc{F}_{T})$, we use that: $|y^{n}|_{S^{\infty}} \vee |Y|_{S^{\infty}} \le M$, to get
\begin{equation}\label{eq: controleA}
 |A_{T}^{n}| \le 2 M + C\disp{\int_{0}^{T} |z_{s}^{n}|^{2}ds + \big|\disp{\sup_{0 \le t \le T}\int_{0}^{t}z_{s}^{n}dW_{s}}\big|}.\end{equation}
Relying on the BDG inequality in $\mc{H}^{p}$ for the last term in (\ref{eq: controleA}), there exists a new constant always denoted by $C$ such that
$$\mb{E}\big(|A_{T}^{n}|^{p}\big) \le C\left(1 +\mb{E}\big(\int_{0}^{T} |z_{s}^{n}|^{2}ds \big)^{p} \right). $$
It remains to show that: $\disp{\sup_{n}\mb{E}\left(\int_{0}^{T} |z_{s}^{n}|^{2}ds \right)^{p}} < \infty$,
and, to achieve this, we first apply It\^o's formula to $e^{\alpha y^{n}}$ 
\begin{tabbing}
  $e^{\alpha y_{t}^{n}} + \disp{\frac{\alpha^{2}}{2}\int_{t}^{T}e^{\alpha y_{s}^{n}}|z_{s}^{n}|^{2}ds } = $ \= $ \disp{e^{\alpha Y_{T}} + \int_{t}^{T}\alpha e^{\alpha y_{s}^{n}}g(s, z_{s}^{n})ds },  $\\
\> $ \quad - \;\disp{\int_{t}^{T}\alpha e^{\alpha y_{s}^{n}}dA_{s}^{n} - \int_{t}^{T}\alpha e^{\alpha y_{s}^{n}}z_{s}^{n}dW_{s}}.   $\\
\end{tabbing}
The next step consists in taking this equation to the power $p$ and then the expectation: this yields the existence of $C$ (depending only on $p$, $\alpha$, $T$ and $|y^{n}|_{S^{\infty}}$) such that
$$ \mb{E}\left|e^{\alpha y_{t}^{n}} + \int_{t}^{T}\alpha e^{\alpha y_{s}^{n}}dA_{s}^{n} + \frac{\alpha^{2}}{2}\int_{0}^{T}e^{\alpha y_{s}^{n}}|z_{s}^{n}|^{2}ds \right|^{p} $$
$$\quad \quad \le \; C\left(\mb{E}\left|\int_{0}^{T}(1 + |z_{s}^{n}|^{2})ds \right|^{p}  + \disp{\mb{E}\big|\sup_{t} \int_{t}^{T}\alpha e^{\alpha y_{s}^{n}}z_{s}^{n}dW_{s}\big|}^{p} \right),$$
To obtain the right-hand side in the previous inequality, we rely both on the assumption ($H_{1}$) on $g^{n}$ and on the boundedness of $Y_{T}$ and we then argue that the left-hand side is greater than
$\mb{E}\left(\frac{\alpha^{2}}{2}\int_{0}^{T}e^{\alpha y_{s}^{n}}|z_{s}^{n}|^{2}ds \big|^{p}\right)$. If we fix $\alpha$ large enough (i.e. $\frac{\alpha^{2}}{2} -  C \ge 1$), we rely on the boundedness of ($y^{n}$) in $S^{\infty}$ and on the BDG inequality for $\disp{\sup_{t}\big|\int_{t}^{T}\alpha e^{\alpha y_{s}^{n}}z_{s}^{n}dW_{s} \big|}$ to claim 
\\
$$\exists \; C >0, \; \textrm{s.t.} \quad  \disp{\mb{E}\left(\int_{0}^{T}|z_{s}^{n}|^{2}ds\right)^{p} \le C + \frac{1}{2}\mb{E}\big(\int_{0}^{T}|z_{s}^{n}|^{2}ds\big)^{p}},$$
which is the desired result.\\

\subsubsection*{Step 3: Convergence results}
In this step, we justify the passage to the limit in the penalized BSDEs with parameters ($g^{n},\; B$)
\begin{equation}\label{eq: equationpenalized}
y_{t}^{n} = B + \disp{\int_{t}^{T}g^{n}(s, z_{s}^{n})ds - \int_{t}^{T}z_{s}^{n}dW_{s}}.  
\end{equation}
To this end, we prove strong convergence results for both ($y^{n} $), ($z^{n}$) and ($A^{n}$).\\
$ \bullet$ From step 2, we first get: $\mb{E}(|A_{T}^{n}|) < \infty$, implying
\begin{equation}\label{eq: convunif} \left( \disp{\mb{E}(\int_{0}^{T}|y_{s}^{n} - Y_{s}|ds)}  \le \frac{\mb{E}(|A_{T}^{n}|)}{n}\right)
\; \textrm{and} \; \left( \frac{\mb{E}(|A_{T}^{n}|)}{n}
\to 0\right),\end{equation}
 since ($A_{T}^{n}$) is bounded in $L^{1}(\mc{F}_{T})$. Thanks to Dini's theorem applied to the decreasing and bounded sequence ($y^{n}$)$_{n}$, both the sequences $\left(\disp{\sup_{n}|y_{s}^{n} - Y_{s}|}\right)$ and $\left(\disp{\sup_{m \ge n}(y_{s}^{n,m})}\right):= \left(\disp{\sup_{m \ge n }|y_{s}^{n} -y_{s}^{ m}|}\right)$ converges to zero, as $n$ goes to $\infty$.\\
$ \bullet$ To justify that ($z^{n}$) is a Cauchy sequence in $\mc{H}^{2}$,
we apply It\^o's formula to $|y^{n, m}|^{2}$ 
\[ \begin{array}{l}
\disp{\mb{E}\left(\int_{0}^{T}|z_{s}^{n, m}|^{2}ds\right)} 
\\
 \le \disp{\mb{E}\left(|y_{0}^{n, m}|^{2} \right)} 
 \; + \;\disp{ 2\mb{E}\left(\int_{0}^{T} |y_{s}^{n, m}|\left(|g(s, z_{s}^{n}) - g(s, z_{s}^{m})|ds + dA_{s}^{n} + dA_{s}^{m}\right)  \right)}
\\
\\
\quad  \le \mb{E}(\disp{\sup_{t} |y_{t}^{n, m}|^{2}}) +\;\disp{2\left( \mb{E}\left(\disp{\sup_{t} |y_{t}^{n, m}|^{2}}\right)\mb{E} \left(\int_{0}^{T}C(1 + |z_{s}^{n}|^{2} + |z_{s}^{m}|^{2})ds + A_{T}^{m} + A_{T}^{n} \right)^{2} \right)^{\frac{1}{2}} }  \\
\\
\quad
\le \mb{E}(\disp{\sup_{t} |y_{t}^{n, m}|^{2}}) + C\mb{E}\big(\disp{\sup_{t} |y_{t}^{n, m}|^{2}})\big)^{\frac{1}{2}},  \\
\end{array} \]
where the last constant $C$ depends only on the estimates of ($z^{n}$) in $\mc{H}^{4}$ and those of ($A^{n}$) in $L^{2}(\mc{F}_{T})$ (for these estimates, we refer here to Step 2). $(z^{n})$ being a Cauchy sequence, it converges in $\mc{H}^{2}$. 
Then, referring to Lemma 2.5 in \cite{mkobylanski}, we argue the existence of $\tilde{z}$ such that, at least along a sequence of integers, 
\begin{equation}\label{eq: integrability}                                                                                  \tilde{z} :=\disp{\sup_{m}|z^{m}|^{2}} \in  \mc{H}^{2} .                                                           \end{equation}
$ \bullet$ To conclude, it suffices to show that ($A_{t}^{n}$) converges in $L^{1}(\mc{F}_{t})$ for all $t$.\\
 We first claim that, between $0$ and $T$ and for any $n , m$, $y^{n, m}$ solves
\begin{equation}\label{eq: convergencecauchy}
y_{0}^{n,m} = \disp{ \int_{0}^{T}(g(s, z_{s}^{n}) - g(s, z_{s}^{m}))ds - \int_{0}^{T}z_{s}^{n ,m}dW_{s} -\big( A_{T}^{n} - A_{T}^{m}\big)}.
 \end{equation}
Now and for any $n,\; m$ such that: $n \le m$, we introduce $g^{n, m}$ as follows $$g^{n, m} := \big( g(s, z_{s}^{n}) - g(s, z_{s}^{m})\big),$$ and we prove that ($g^{n, m}$) is a Cauchy sequence in $L^{2}([0, T], \mc{F})$ and hence, strongly convergent in $L^{2}([0, T], \mc{F})$).
Relying on assumption ($H_{1}$), we obtain 
$$ 
\exists \; \lambda \in BMO(W), \quad \; \disp{
| g(s, z_{s}^{n} - g(s, z_{s}^{m})| \le |\lambda_{s}(z_{s}^{n}, \;z_{s}^{ m})|
|z_{s}^{n ,m}|}.$$ 
Both assumption ($H_{1}$) and the strong convergence of ($z^{n}$) in $\mc{H}^{2}$ yields that $\lambda^{n, m} := (\lambda_{s}(z_{s}^{n}, \;z_{s}^{ m}))$ is dominated uniformly in $n$ and $m$ by $ C(\kappa + |\tilde{z}|)
$, which is an integrable variable (thanks to (\ref{eq: integrability})).
Using that $\tilde{z}$ and $\kappa$ are in BMO($W$), we obtain that $\lambda:= \disp{\sup_{n, m} \lambda^{n, m}} $ is itself in BMO($W$). Hence, duality between the space of BMO martingales and $\mc{H}^{2}$ entails 
$$ \exists \; C > 0,\quad 
\disp{\mb{E}\big(\int_{0}^{T}| g(s, z_{s}^{n} - g(s, z_{s}^{m})|ds\big)} \le  C|\lambda|_{\textrm{BMO}} \disp{\mb{E}\left(\int_{0}^{T}|z_{s}^{n, m}|^{2}ds\right)^{\frac{1}{2}}}.$$
Since $(z^{n, m})$ is a Cauchy sequence, this implies that $(g^{n, m})$ is itself a Cauchy sequence.\\
We now rewrite equation (\ref{eq: convergencecauchy}) between $0$ and $t$, which gives
$$y_{0}^{n,m} -y_{t}^{n, m}  = \disp{ \int_{0}^{t}g^{n, m}ds - \int_{0}^{t}z_{s}^{n ,m}dW_{s} -\big( A_{t}^{n} - A_{t}^{m}\big)},
 $$
 and we transfer the last term $A_{t}^{n, m}$ into the left-hand side. Taking then successively the absolute value, the supremum over $t$ and the expectation, we obtain
$$ \disp{\mb{E}
\big(\disp{\sup_{t}| A_{t}^{n} - A_{t}^{m} |} \big) \le 2\mb{E}\left(\disp{\sup_{t}|y_{t}^{n, m}|} +\disp{\sup_{t}|\int_{0}^{t}z_{s}^{n}dW_{s}|} + \int_{0}^{T}|g(s, z_{s}^{n} - g(s, z_{s}^{m})|ds  \right) }.$$
We next rely on the BDG inequality for the square integrable martingales $z^{n,\;m} \cdot W$ and on the previous results to conclude that, for all $t$, the sequence of processes $(A_{\cdot}^{n})$ is Cauchy in $L^{1}\big([0, T],\; \mc{F} \big)$: we denote by $K$ its limit, which is increasing as limit of such processes and we denote by $z$ the limit of ($z^{n}$) in $\mc{H}^{2}$. Passing to the limit in (\ref{eq: equationpenalized}) as $n$ goes to $\infty$, we get 
$$ Y_{t} := Y_{T} +\disp{\int_{t}^{T}g(s, z_{s})ds -(K_{T} - K_{t})  - \int_{t}^{T}z_{s}dW_{s} },$$
which is the desired decomposition of the $g$-submartingale $Y$.
\begin{flushright}
 $\square$
\end{flushright}

\subsection{Characterization of the solution of the RBSDE}
To justify the expression of the solution given in Theorem \ref{SolutionRBSDE}, we rely both on the results of the previous section and on the characterizations already provided in Proposition 2.3 and Proposition 5.1 in \cite{ElkarouiKapoudian}. In this paper, the authors prove that
the solution ($Y, Z, K$) of a RBSDE with driver $f:=f(s,y, z)$, lower obstacle $S$ and terminal condition $\xi$ satisfies
\begin{equation}\label{eq: RBSDE}
 Y_{t} := \disp{ \textrm{ess} \sup_{\tau \in \mc{S}_{t, T}} \mb{E}\left( \int_{t}^{T}f(s, Y_{s}, Z_{s})ds +\xi + S_{\tau}\mathbf{1}_{\tau \le T} | \mc{F}_{t} \right)},\end{equation}
where $\mc{S}_{t, T}$ stands for the set of all stopping times taking values in $[t, T]$.
Here, contrary to the aforementionned paper, where the generator $f:=f(s,y, z)$ of the RBSDE is assumed to be lipschitz both in $y$ and $z$, we relax this last assumption.\\ 
Hence, to characterize the solution of the RBSDE by a formula similar to (\ref{eq: RBSDE}), we need the extension of the Doob-Meyer's decomposition for non linear $g$ expectations (this last one has been obtained in Section 2.4): let $\tilde{Y}$ be equal to
\begin{equation}\label{eq: expresssolution} 
\tilde{Y}_{t} := \textrm{ess} \disp{\inf_{\tau \in \mc{S}_{t, T}} \mc{E}_{g}\big(B\mathbf{1}_{\tau = T} +U_{\tau}\mathbf{1}_{\tau < T} +\int_{t}^{\tau}f_{0}(s, 0)ds |\mc{F}_{t} \big)},\end{equation}
with $g$ satisfying both $(H_{0})$ and $(H_{1})$ and such that: $g(s, 0) \equiv 0$.
Our aim is to prove that such a process can be characterized as the largest $g$-submartingale dominated by the upper obstacle $U$ and hence that it solves the equation $ (\textrm{Eq2.1})$. To this end, we proceed by justifying the two following arguments:\\
$\bullet \;$ the $g$-submartingale property of $\tilde{Y}$,\\
 $\bullet \;$  the optimality among the class of $g$ submartingales (smaller than $U$).\\

\paragraph*{Step 1: Submartingale property}
We consider an arbitrary pair $s, t$ such that: $s \le t$. We aim at proving that the process $ \tilde{Y}$ given by (\ref{eq: expresssolution}) satisfies: $\tilde{Y}_{s} \le \mc{E}_{g}\big(\tilde{Y}_{t}| \mc{F}_{s}\big)$.
For this and for an arbitrary stopping time $\tau$, we set $H_{\cdot, \tau}$ as follows $$H_{t, \tau}:=\disp{ B\mathbf{1}_{\tau = T} +U_{\tau}\mathbf{1}_{\tau < T} +\int_{t}^{\tau}f_{0}(s, 0)ds}.$$ 
Since: $\mc{S}_{t, T} \subset \mc{S}_{s, T}$, the essential infimum taken over the subset $\mc{S}_{t, T}$ is then strictly greater than the one taken over $\mc{S}_{s, T}$: hence,
$$ \tilde{Y}_{s} \le \textrm{ess} \disp{\inf_{\tau \in \mc{S}_{t, T}} \mc{E}_{g}\big(H_{s, \tau}|\mc{F}_{s} \big)}$$
Then, using the inequality: $f_{0} \ge 0$, we check that, for any $s, t,$ $s \le t$ and any $\tau$ in $\mc{S}_{t, T}$: $H_{s, \tau} \le H_{t, \tau} $. This yields
\[  \begin{array}{ll} 
  \tilde{Y}_{s}  &\le \textrm{ess} \disp{\inf_{\tau \in \mc{S}_{t, T}} \mc{E}_{g}\big(H_{t, \tau}|\mc{F}_{s} \big)},\\
   & \le \textrm{ess} \disp{\inf_{\tau \in \mc{S}_{t, T}} \mc{E}_{g}\left(\mc{E}_{g}\big(H_{t, \tau}|\mc{F}_{t} \big)  |\mc{F}_{s} \right)}.\\
\end{array} \]
The last part of the proof consists in justifying that we can reverse the roles of the essential infimum and of the conditional $g$ expectation $\mc{E}_{g}(\cdot | \mc{F}_{s})$, which means
\begin{equation}\label{eq: interversion}
\textrm{ess} \disp{\inf_{\tau \in \mc{S}_{t, T}} \mc{E}_{g}\left( \mc{E}_{g}\big(H_{t, \tau}|\mc{F}_{t} \big)|\mc{F}_{s} \right)} = \mc{E}_{g}\left( \disp{\left(\textrm{ess} \inf_{\tau \in \mc{S}_{t, T}} \mc{E}_{g}\big(H_{t, \tau}|\mc{F}_{t} \big) \right) |\mc{F}_{s} }  \right)\; ,\end{equation}
where the right hand-side member coincide with $\mc{E}_{g}\big( \tilde{Y}_{t} |\mc{F}_{s} \big)$.
To obtain a first inequality, we rely on the Fatou property for the conditional $g$-expectation $ \mc{E}_{g}( \cdot|\mc{F}_{s})$ to claim
$$ \textrm{ess} \disp{\inf_{\tau \in \mc{S}_{t, T}} \mc{E}_{g}\left(\mc{E}_{g}\big(H_{t, \tau}|\mc{F}_{t} \big)|\mc{F}_{s} \right)} \ge \mc{E}_{g}\left( \textrm{ess} \disp{\inf_{\tau \in \mc{S}_{t, T}} \mc{E}_{g}\big(H_{t, \tau}|\mc{F}_{t} \big) } |\mc{F}_{s} \right).$$
For the other inequality, we consider a minimizing sequence $(\tau^{n})$ of stopping times in $\mc{S}_{t, T}$ such that
\begin{equation}\label{eq: convergence} 
\mc{E}_{g}\big(H_{t, \tau^{n}}|\mc{F}_{t} \big) \rightarrow \textrm{ess} \disp{\inf_{\tau \in \mc{S}_{t, T}} \mc{E}_{g}\big(H_{t, \tau}|\mc{F}_{t} \big)}, \quad \textrm{as} \;n \; \to \infty. 
\end{equation} 
Such a sequence ($\tau^{n}$) exists, since the family $(Z_{t, \tau}) =( \mc{E}_{g}\big(H_{t, \tau}|\mc{F}_{t} \big))_{ \tau \;\in \; \mc{S}_{t, T}}$ is stable by taking the infimum: in fact, one can check
$$\mc{E}_{g}\big(H_{t, \tau^{1}}|\mc{F}_{t} \big) \wedge \mc{E}_{g}\big(H_{t, \tau^{2}}|\mc{F}_{t} \big) = \mc{E}_{g}\big(H_{t, \tau^{*}}|\mc{F}_{t} \big), $$
where the stopping time $\tau^{*}$ is defined as follows:
\[\left\{ \begin{array}{ll}
 \tau^{*} := \tau^{1}, \quad & \textrm{if}\; \omega \in \{Z_{t, \tau^{1}} \le Z_{t, \tau^{2}}  \},   \\
\\
   \tau^{*} := \tau^{2} \quad & \textrm{else}. \\
\end{array} \right. \]
Without loss of generality, we assume that the convergence in (\ref{eq: convergence}) is decreasing. Hence, using that such conditional quadratic $g$-expectations satisfy a stability result (for a precise statement of this result, we refer to Proposition 2.4 in \cite{mkobylanski}), it yields
$$\mc{E}_{g}\big(\mc{E}_{g}\big(H_{t, \tau^{n}}|\mc{F}_{t}\big) |\mc{F}_{s}\big) \to  \mc{E}_{g}\left( \textrm{ess} \disp{\inf_{\tau \in \mc{S}_{t, T}} \mc{E}_{g}\big(H_{t, \tau}|\mc{F}_{t} \big) } |\mc{F}_{s} \right). $$
To conclude, we argue that $\mc{E}_{g}\big(\mc{E}_{g}\big(H_{t, \tau^{n}}|\mc{F}_{t}\big) |\mc{F}_{s}\big) := \mc{E}_{g}\big(H_{t, \tau^{n}}|\mc{F}_{s}\big) $: as a consequence, its limit, as $n$ goes to $\infty$, is greater than the left hand side of the equality (\ref{eq: interversion}), which yields the second inequality and ends the proof of this step.\\ 
\paragraph*{Step 2: Optimality}
To achieve the proof of the optimality, we just need to show that the solution $\tilde{Y}$ given by (\ref{eq: expresssolution}) satisfies the condition \begin{equation}\label{eq: saturation}
\disp{\int_{0}^{T}(\tilde{Y}_{s} - U_{s})d\bar{K}_{s}} = 0,
 \end{equation}
for a well chosen increasing process $\bar{K}$.
For this, we fix $t$ and we introduce the stopping time $D_{t}$
$$ D_{t} := \disp{ \inf \{u, \; u \ge t, \; \tilde{Y}_{u} = U_{u}  \}} \wedge T.$$
By convention, $ \disp{\inf \{\emptyset\}} = \infty$. As soon as $D_{t} < T$, we get: $\tilde{Y}_{D_{t}} = U_{D_{t}}$, and defining $\bar{Y}$ by: $\forall \;t, \quad \bar{Y}_{t} = \tilde{Y}_{t} + \disp{\int_{0}^{t}f_{0}(0, s)ds} $, this implies \begin{equation}\label{eq: optimaleeq} \bar{Y}_{t} := \mc{E}_{g}\big(\bar{Y}_{D_{t}} |\mc{F}_{t} \big). \end{equation}
Since $\tilde{Y}$ is a $g$-submartingale, there exists a $g$-martingale $M$ and an increasing process $\tilde{K} $ such that: $\tilde{Y} = M + \tilde{K}$.
 It follows that
$$ \bar{Y}_{D_{t}} - \bar{Y}_{t}  = (M_{D_{t}} - M_{t}) + (\tilde{K}_{D_{t}} - \tilde{K}_{t}) +\disp{\int_{t}^{D_{t}}f_{0}(0, s)ds}.$$
We introduce $\bar{K}$ as: $\bar{K}:= \tilde{K} + \disp{\int_{0}^{\cdot}f_{0}(s, 0)ds }$, which is an increasing process and we take the conditional $g$-expectation $\mc{E}_{g}\big(\cdot |\mc{F}_{t} \big)$ in both sides of the previous equality using (\ref{eq: optimaleeq}): this yields 
\[ \begin{array}{ll}
 
0 =  \mc{E}_{g}\big( \bar{Y}_{D_{t}} - \bar{Y}_{t}|\mc{F}_{t} \big) & =   \mc{E}_{g}\big( (M_{D_{t}} - M_{t}) + (\bar{K}_{D_{t}} - \bar{K}_{t}) |\mc{F}_{t} \big)\\
\\
    &  \ge  \mc{E}_{g}\big( M_{D_{t}} - M_{t}|\mc{F}_{t} \big) =0.\\ 
\end{array} \]
To justify that: $\mc{E}_{g}\big( \bar{Y}_{D_{t}} - \bar{Y}_{t}|\mc{F}_{t} \big) =0 $, we use the invariance by translation property of $ \mc{E}_{g}(\cdot |\mc{F}_{t})$ (the same equality holds if we replace $\bar{Y} $ by $M$). Then, to prove the inequality, we use both the monotonicity of $ \mc{E}_{g}(\cdot |\mc{F}_{t}) $ and the increasing property of the process $\bar{K}$. All inequalities being equalities, we finally get: $\bar{K}_{t} = \bar{K}_{D_{t}}.$ This means that, on the set $\{\tilde{Y} < U \} $, the increasing process $\bar{K}$ is constant, which yields (\ref{eq: saturation}) and ends the proof.
\begin{flushright}
 $\square$
\end{flushright}

\section{Representation of the solution as a non linear time consistent pricing rule}
Our aim is to show that the solution of the RBSDE with driver $g$ and upper bound the american contingent claim $H$, which is given by
\begin{equation}\label{eq: solutionRBSDE} Y_{t} = \textrm{ess} \;\disp{\sup_{\tau \in  \mc{S}_{t, T}} \mc{E}_{g}(H |\mc{F}_{\tau})}, \end{equation}
extends the particular dynamic concave monetary functional introduced in a discrete time setting in Section 5.3
of the paper \cite{cheriditodelbaen} . Here, setting: $g_{\alpha}(s, z) := \frac{\alpha}{2}|z|^{2}$ and as soon as the (normalized) driver $g$ is dominated by $ g_{\alpha}$, the use of the comparison theorem entails that: $\mc{E}_{g}(H |\mc{F}_{\tau}) \le \mc{E}_{g_{\alpha}}(H |\mc{F}_{\tau}). $\\ \\
In the aforementionned paper \cite{cheriditodelbaen}, where the authors work on a discrete time setting, they introduce a functional, referred as the dynamic concave monetary functional defined by worst stopping: this functional maps $L^{\infty}(\mc{F}_{T})$ in $L^{\infty}(\mc{F}_{t}) $ as follows
 \begin{equation}\label{eq: worststoppfunctional} \Psi_{t, T}(\xi) := \textrm{ess} \disp{\inf_{\tau \in\mc{S}_{t, T}}\Phi_{\tau, T}(\xi_{\tau}) },
 \end{equation}
where $\tau$ runs over all stopping times taking values in $[t, T]$ and $\Phi_{\tau, T} $ is an arbitrary dynamic (concave) monetary functional: it maps $L^{\infty}(\mc{F}_{T})$ into $L^{\infty}(\mc{F}_{\tau})$. Here, this definition makes sense if and only if $\xi_{\tau}$ refers to a $\mc{F}_{\tau }$-measurable random variable: for instance, when $\xi$ denotes an $\mc{F}_{T}$-random variable, it can be: $\mc{E}(\xi |\mc{F}_{\tau})$.\\
We now consider the special case of the dynamic entropic risk functional denoted by $\rho^{\alpha}$ and introduced at the end of section 1.1: it is given by $$ \forall X \in L^{\infty}(\mc{F}_{T}), \quad \rho^{\alpha}_{t}(X) = \frac{1}{\alpha}\ln \mb{E}(e^{-\alpha X} | \mc{F}_{t}).$$
 (in the context of dynamic functionals defined on a finite time horizon $T$, $\rho^{\alpha}_{t}$ stands for $\rho^{\alpha}_{t, T}$). Introducing now $\Phi^{\alpha}$ by setting 
$$\forall \; X \in L^{\infty}(\mc{F}_{T}), \quad \Phi^{\alpha}_{\tau, T}(X) = -\rho_{\tau, T}^{\alpha}(X). $$
 we obtain a dynamic (concave) monetary functional.\\
Referring to the comments given at the end of section 1.1, we deduce 
$$\forall \; X \in L^{\infty}(\mc{F}_{T}), \quad  -\Phi^{\alpha}_{\tau, T}(-X) = \mc{E}_{g_{\alpha}}(X |\mc{F}_{\tau}) = \rho_{\tau, T}^{\alpha}(-X) , $$
where $g_{\alpha}(z) := \frac{\alpha}{2}|z|^{2}$. If $\Psi^{\alpha}$ is defined in terms of $ \Phi^{\alpha}$ as in (\ref{eq: worststoppfunctional}), then it follows  
\[\begin{array}{ll}
- \Psi^{\alpha}_{t, T}(-H) &= -\textrm{ess}\disp{\inf_{\tau \in\mc{S}_{t, T}}\Phi^{\alpha}_{\tau, T}(-H)} =\textrm{ess}\disp{\sup_{\tau \in\mc{S}_{t, T}}-\Phi^{\alpha}_{\tau, T}(-H)} \\
\\
& = \textrm{ess} \disp{\sup_{\tau \in\mc{S}_{t, T}}\mc{E}_{g_{\alpha}}(X |\mc{F}_{\tau})}.\\
\end{array} \]
Hence, the process given by (\ref{eq: solutionRBSDE}) satisfies $$Y_{t} \le \textrm{ess} \;\disp{\sup_{\tau \in  \mc{S}_{t, T}} \mc{E}_{g_{\alpha}}(H |\mc{F}_{\tau})} = - \Psi^{\alpha}_{t, T}(-H).$$  \\

\subsubsection*{Comments}
\textbullet  $\;$
 The solution of the quadratic RBSDE can be reinterpreted a valuation formula for the American contingent claim: in fact, this valuation formula is analogous to the one given in \cite{ElkarQuenez} in which the authors define the price of the American option as being the upper price of European type options. Our characterization generalizes the representation as a Snell envelope by introducing the extented notion of $g$-Snell envelope ($g$ being a quadratic convex driver).\\
\textbullet $\;$
A second comment is that there exists an interpretation via convex duality theory:
this is already given in \cite{ElKaretRouge}, where the authors relate this price with the exponential utility maximization problem and using dual formulations. Dual formulation leads also more generally to the robust representation of the dynamic concave utility functional such as the one denoted by $\Phi^{\alpha}$ in this paragraph (such functionals are defined in a more general setting in \cite{KloppelSchweizer}).
\newpage
\subsection{One financial application}
As an example, we give the description of the problem in \cite{Cvitanicwan07} and explain both the origin and interpretation of the solution of the RBSDE they introduce.\\
In that paper, the authors adress the problem of defining a specific notion of optimal contract and they assume that the contract can be exerced at any random time in the context of a continuous-time double agent problem. 
The filtration considered is a brownian one and, for any process $u$ such that the exponential of the stochastic integral $\disp{\int_{0} u_{s}dB_{s}}$ is a true martingale \footnote{This condition is checked in particular under Novikov's condition \begin{equation}\label{eq: conditionintegrab} \mb{E}\left(\exp\big( \frac{1}{2}\int_{0}^{T}|u_{s}|^{2}ds \big) \right) < \infty \end{equation} or when the stochastic integral of $u$ w.r.t $B$ is in the class of BMO martingales.},
the notation $\mb{P}^{u}$ stands for the equivalent measure defined by
$$\frac{d\mb{P}^{u}}{d\mb{P}} = \exp\big(\int_{0}^{t}u_{s}dB_{s} - \int_{0}^{t}\frac{1}{2}|u_{s}|^{2}ds \big) = \mc{E}( u \cdot B),$$
and the output controlled process $X^{u}$ is such that
\begin{equation}
dX^{u} = u_{t}v_{t}dt + v_{t}dB^{u}, 
\end{equation}
where $u$ stands for the control process and $B^{u}$ is the brownian motion obtained by the usual Girsanov's transform under $\mb{P}^{u}$.
 If we consider a given contract, i.e. a family of $\mc{F}_{t}$-measurable random variables $C$ = ($C_{\tau}$) standing for the random payment and if we assume that the agent has the ability to choose the exercise time, then the formulation of the agent's problem having $U$ for utility function is 
\begin{equation}\label{eq: controlproblem}
 \disp{\sup_{u, \tau}\mb{E}^{u} \left( U(\tau, C_{\tau}) - \int_{t}^{\tau}g(u_{s})ds |\mc{F}_{t}\right)},
\end{equation}
where the notation $\mb{E}^{u}$ stands for the expectation under $\mb{P}^{u}$
and the integral $\disp{\int_{0}^{t}g(u_{s})ds}$ describes the cumulative cost, which is due to early exercise of the contract and that the agent has to pay.
Denoting as usual by $I$ the functional given in terms of $U$ by: $I(\cdot) = (U^{'})^{-1}$ \footnote{The functional $I$ is also introduced in the optimization problem studied in \cite{kramkovschach} and it is used in the characterization of the optimal strategy in terms of the solution of the dual control problem, which is the density of one martingale measure.},
 the solution of (\ref{eq: controlproblem}) is characterized as the unique process $W^{A} $ solving the RBSDE
\[ \left\{ \begin{array}{ll}
W_{t}^{A} = U(t, C_{t}) & + \disp{\int_{t}^{T} \big(g(I(w_{s}^{A})) - w_{s}^{A}I(w_{s}^{A}) \big)ds - \int_{t}^{T}w_{s}dB_{s} - (K_{T}^{A} - K_{t}^{A})}\\
\\
W_{t}^{A} \ge U(t, C_{t}) & \textrm{and}\; \disp{\int_{0}^{T}\big(W_{t}^{A} - U(t, C_{t})\big) dK_{t}^{A}}=0.   \\
\end{array} \right.  \]
Referring to the same arguments in \cite{ElkarouiKapoudian}, the optimal exercise time $\tau^{A}$ is: $\tau^{A} = \disp{\inf \{ t, \; W_{t}^{A} = U(t, C_{t}) \}}$,
and the optimal control $u = u^{A}$ is given in feedback form by: $\forall \; s, \;\;u_{s}^{A} = (U^{'})^{- 1}(w_{s}^{A}).$
Hence, if we rewrite the backward equation under $\mb{P}^{u^{A}}$ between $t$ and $\tau^{A}$, it
 yields $$ W_{t}^{A} = \mb{E}_{t}^{u_{A}}\left(U(\tau^{A}, C_{\tau^{A}}) - \disp{\int_{t}^{\tau^{A}} \big(g(u_{s}^{A}) \big) ds} \right), $$
 which implies the optimality in (\ref{eq: controlproblem}) of the control $u^{A}$ associated to the equivalent probability measure $\mb{P}^{u^{A}} $.\\

\subsubsection*{Comments}

$\bullet \;$ Analogously as in subsection 3.1 and using the expression (\ref{eq: controlproblem}) of the problem, it follows that the solution $W$ of the RBSDE is related to the following DMCUF
\begin{equation}\label{eq: DMCUF}
 u_{t}(Y) = \disp{\textrm{ess}\inf_{u, \tau} \mb{E}^{\mb{P}^{u}}\left( Y_{\tau}  + \alpha_{t, \tau}(\mb{P}^{u}) | \mc{F}_{t}\right) },
\end{equation}
with $\alpha_{t, \tau}(\mb{P}^{u}) = \disp{\int_{t}^{\tau^{A}} g(u_{s}^{A}) ds}$.
The relation between these two processes is given for all $t$ by 
 $$ \;  W_{t}^{A} = u_{t}(-Y),\quad  \textrm{with} \;Y \; \textrm{such that}:  Y_{\cdot} = U(\cdot, C_{\cdot}). $$
Besides, relation (\ref{eq: DMCUF}) coincide with the robust representation of the dynamic utility functional $u = (u_{t})_{t}$ (or, up to a sign, to the related dynamic risk functional $\rho = -u $): in this brownian setting, the set of martingale measures is given by the family $ \{ \mc{E}(u \cdot B) \; \textrm{such that} \; (\ref{eq: conditionintegrab}) \; \textrm{holds}\}$.\\
$\bullet \;$ One major restriction 
in this example is that the agent can only act on the drift of the output process $X$
and hence, modulo a penalized term given explicitely in terms of the cost function $g$, the optimization problem reduces to a control problem over the set of measures $\mb{P}^{u}$.\\
$\bullet \;$ The more general control problem associated with the utility maximization problem with random time horizon and utility function $U$
 is discussed in the complete case in \cite{Karatzas}. Even in that case, the characterization of optimal stopping time and optimal strategies in the following problem
$$V = \disp{\sup_{\theta, \; \tau} \mb{E} \big( U(X_{\tau}^{\theta})\big)}, $$
where $ X^{\theta}$ stands for the wealth process obtained by using an admissible strategy $\theta$,
is not trivial. It is even proved that optimal strategies may not exist in general.

\end{document}